\documentstyle[preprint,prl,aps]{revtex}

\def\pmb#1{\setbox0=\hbox{$#1$}%
\kern-.025em\copy0\kern-\wd0
\kern.05em\copy0\kern-\wd0
\kern-.025em\raise.0433em\box0}

\def\beq{\begin{equation}}
\def\eeq{\end{equation}}

\begin{document}

\tighten 

\def\footnoterule{\hrule width \hsize}
\def\footstrut{\baselineskip 16pt}

\skip\footins = 14pt 
\footskip     = 20pt 
\footnotesep  = 12pt 

\title{%
REDUCING THE CHERN--SIMONS TERM \\ BY A SYMMETRY\footnotemark[1]
}

\footnotetext[1] {\baselineskip=16pt This work is supported in part by funds
provided by  the U.S.~Department of Energy (D.O.E.) under contract
\#DE-FC02-94ER40818 and under contract \#DE-FG02-91ER40676. \newline
MIT-CTP-2696 \qquad BU-HEP-97-30\qquad hep-th/9712087 \hfil  November
1997\break}

\author{R.~Jackiw}

\address{Center for Theoretical Physics\\ Massachusetts Institute of
Technology\\ Cambridge, MA ~02139--4307}

\author{S.-Y. Pi}

\address{Physics Department\\ Boston University \\ Boston, MA ~02215}

\maketitle

\thispagestyle{empty}

\begin{abstract}\noindent
Reducing a 3-dimensional Chern--Simons term by a symmetry yields other
topologically interesting structures.  Specifically, reducing by radial
symmetry results in a 1-dimensional quantum mechanical model, which has
recently been used in an analysis of finite-temperature Chern--Simons theory. 
The radially symmetric expression may be inserted into 3-dimensional monopole
or $(2+1)$-dimensional instanton equations, where it eliminates the monopole
or instanton solutions.
\end{abstract}

\vskip.25in

\centerline{Submitted to {\it Physics Letters B}}

\vskip.33in

The gauge dependence of the induced Chern--Simons term at finite
temperature has been recently explained \cite{ref:1,ref:2}.  Original
calculations seemed to indicate that the magnitude of this topological entity
varies smoothly with temperature \cite{ref:3}, contradicting the
integrality of the Chern--Simons coefficient, as is required in the quantum
theory by gauge invariance\cite{ref:4}.  The puzzle became resolved once it was
realized that finite-temperature calculations to fixed perturbative order
(like the original ones
\cite{ref:3}) necessarily violate gauge invariance, which is restored only
after all orders are summed. (At zero temperature, finite order
calculations suffice to exhibit the complete, induced Chern--Simons
term\cite{ref:4,ref:5}.)

The all-order summation was first accomplished in a toy quantum mechanical
model, which had been introduced a decade earlier for the purpose of
exhibiting in a simple setting some of the peculiar topological/geometrical
effects of quantized Chern--Simons theory\cite{ref:6}.  In this Letter,
we demonstrate that this model is not merely a pedagogical toy; in fact it
coincides with the three-dimensional Chern--Simons term, reduced by spherical
symmetry.

We first review the quantum mechanical analog, and then demonstrate how it
arises by symmetric reduction.  We conclude with remarks about the effect of
the Chern--Simons term on the monopole and instanton equations.

\section{Chern--Simons Quantum Mechanics}

Let us record the Lagrangian of $(2+1)$-dimensional, topologically massive
(Chern--Simons) electrodynamics, in the Weyl ($A_0= 0$) gauge.
\begin{equation}
L_{\rm CS/EM} = \frac12 \int d^2x \, \Bigl(  (\dot{A}_m)^2 -  (\nabla
\times A)^2_m + \mu\epsilon^{mn} A_m \dot{A}_n \Bigr)
\label{eq:1}
\end{equation}
The over-dot indicates time-differentiation of the two-component field
variable $A_m(t,{\bf x})$, $(m=1,2)$.  The above expression suggests the
following quantum mechanical analog, which involves a two-component particle
variable $q^m(t)$
\begin{equation}
L_{\rm QM} = \frac12 \Bigl( (\dot{q}^m)^2 -  \omega^2 (q^m)^2 +
\mu \epsilon^{mn} q^m \dot{q}^n \Bigr)
\label{eq:2}
\end{equation}
The similarity of respective structures in (\ref{eq:1}) and (\ref{eq:2}) is
self-evident.  $L_{\rm QM}$ describes a particle on the plane, moving in a
constant magnetic field $\mu$ perpendicular to the plane; this is a ``field
theory'' in one dimension, time.  Note that the theory is invariant against
global rotation by time-independent angle
$\theta$.
\begin{equation}
\delta q^m = \epsilon^{mn}q^n \theta
\label{eq:3}
\end{equation}
The corresponding constant of motion is the angular momentum
\begin{equation}
M= \epsilon^{mn}q^m p_n
\label{eq:4}
\end{equation}
whose value is not fixed, but  restricted to
integers in the quantum theory (when the configuration-space plane has no
punctures).

In spite of the similarity of (\ref{eq:1}) to (\ref{eq:2}), the latter, unlike
the former, is not a gauge theory: the $U(1)$ symmetry (\ref{eq:3}) of $L_{\rm
QM}$ is global.  But the invariance can be localized and $\theta$ can be an
arbitrary function of time, provided a connection $a$ is introduced and the
time derivative of $q^m$ is replaced by a gauge covariant derivative
\begin{equation}
\dot{q}^m \to (Dq)^m \equiv \dot{q}^m - \epsilon^{mn}aq^n
\label{eq:new5}
\end{equation}
$(Dq)^m$ transforms covariantly when (\ref{eq:3}) is supplemented by
\begin{equation}
\delta a = \dot{\theta}
\label{eq:5}
\end{equation}

In one (time) dimension, we cannot form a curvature from the connection $a$,
but because the dimensionality is odd we can construct a Chern--Simons term: it
is just $a$.  Thus the model that we adopt as the one-dimensional analog of
$L_{\rm CS/EM}$ is the gauge invariant version of (\ref{eq:2}), supplemented
with a Chern--Simons term of strength $-\nu$.
\begin{equation}
L_{\rm CS/QM} = \frac12 \Bigl(((Dq)^m)^2 - \omega^2 (q^m)^2 +
\mu \epsilon^{mn} q^m (Dq)^n  \Bigr)-\nu a
\label{eq:6}
\end{equation}

Since we may always work in the ``Weyl'' gauge $a=0$, the dynamical equations
of (\ref{eq:6}) coincide with those of (\ref{eq:2}), except that (\ref{eq:6})
also enforces a ``Gauss law,'' which follows upon varying
$a$, and reads
\begin{equation}
M=\nu
\label{eq:7}
\end{equation}
Thus it is seen that the angular momentum, which is not fixed in the un-gauged
theory (\ref{eq:2}), must take the value $\nu$ in the gauged model with
Chern--Simons term (\ref{eq:6}).  Of course in the quantum theory $\nu$ must
be an integer.  This
quantization requirement is alternatively demanded by gauge invariance when
$L_{\rm CS/QM}$ is quantized: under the gauge transformation
(\ref{eq:3}),(\ref{eq:5}), the action $I_{\rm CS/QM}=\int dt L_{\rm CS/QM}$
changes by $\Delta I_{\rm CS/QM}=-\nu \Delta \theta$; where $\Delta \theta =
\int d\theta$; when $\Delta \theta$ is restricted to an integral multiple of
$2\pi$, gauge invariance of $e^{iI_{\rm CS/QM}}$ is assured with integer
$\nu$.  (The situation here is very similar to what happens with Dirac's
monopole, whose quantization also follows alternatively from angular momentum
quantization or from gauge invariance.)

The recent analysis of the Chern--Simons term at finite
temperature\cite{ref:1,ref:2} was based on the model (\ref{eq:6}).  It is clear
that the geometrical/topological effects arise from the last two terms in that
Lagrangian.  We now show how just these terms can be obtained from the
3-dimensional Chern--Simons term.

\section{Reduction of 3-dimensional Chern--Simons Term}

Consider the Chern--Simons action in three space.
\begin{eqnarray}
NW(A) &=& \frac{N}{8\pi} \int d^3x \epsilon^{ijk} (\partial_i A^a_j A^a_k +
\frac13 f_{abc} A^a_iA^b_jA^c_k) \nonumber \\
&=& -\frac{N}{4\pi} \int d^3x \epsilon^{ijk} {\rm tr}(\partial_i A_j A_k +
\frac23 A_iA_jA_k)
\label{eq:8}
\end{eqnarray} 
In the second expression the connections $A_i$ are anti-Hermitian elements of a
Lie algebra, with structure constants $f_{abc}$.  The coefficient is chosen so
that the quantization condition is obeyed with integer $N$.  Let us consider
the $SU(2)$ case and take for $A_i^a$ the radially symmetric {\it Ansatz},
familiar from monopole/instanton studies.
\begin{equation}
A^a_i = (\delta^a_i - \hat{r}^i \hat{r}^a) \frac{1}{r}\psi_1 +
\epsilon^{iaj}\hat{r}^j \frac{1}{r} (\psi_2-1) + \hat{r}^i \hat{r}^a A
\label{eq:9}
\end{equation}
Here $\psi_m$, $m=1,2$, and $A$ are functions just of $r$. Substituting
(\ref{eq:9}) into (\ref{eq:8}), performing the angular integrals, leaves
\begin{equation}
NW(A) = N \int^\infty_0 dr \Bigr(\epsilon^{mn}\psi_m (D\psi)_n-A \Bigl)
 \label{eq:10}
\end{equation}
\begin{equation}
(D\psi)_m \equiv \psi'_m - \epsilon_{mn} A\psi_n
\label{eq:11}
\end{equation} 
Here the dash denotes $r$-differentiation, and we have dropped an endpoint
contribution, $ \psi_1 |^{r=\infty}_{r=0}$. Comparison with (\ref{eq:6}) shows
that the geometric/topological portions of that expression coincide with
(\ref{eq:10}) (except that $r$ has only half the range of $t$).  Moreover the
strength $\nu$ in (\ref{eq:6}), here emerges as the integer $N$, consistent
with the quantization requirement.

The construction may be generalized to an arbitrary group, where the radially
symmetric {\it Ansatz}, which generalizes (\ref{eq:9}), is best presented when
the Lie algebra-valued vector $A_i$ is decomposed into spherical components
$(A_r, A_\theta, A_\phi)$\cite{ref:7}.  One chooses a constant element $\Omega$
lying in the Cartan sub-algebra of the Lie algebra.  Then $A_r$, a function
only of
$r$, is also a member of the Cartan sub-algebra.  The remaining components
$A_\theta$ and $A_\phi$ are given by
\begin{eqnarray}
A_\theta &=& \frac1r \Psi_1 \nonumber \\
A_\phi &=& \frac1r \Psi_2 - \frac1r {\rm ctn} \theta \Omega
\label{eq:12}
\end{eqnarray} 
Here $\Psi_1$ and $\Psi_2$ are functions of $r$ and satisfy
\begin{equation}
[\Psi_m, \Omega] = \epsilon_{mn} \Psi_n
\label{eq:13}
\end{equation} 
(Depending on the choice of $\Omega$, it may be that the only solution  is
$\Psi_m=0$.) Substituting these formulas into (\ref{eq:8}) gives
$N \int^\infty_0 dr {\rm tr} \Bigr\{ \epsilon^{mn} \Psi_m (D\Psi)_n - A_r
\Omega \Bigl\}$ with 
\begin{equation}
(D\Psi)_m \equiv \Psi'_m + [A_r, \Psi_m]
\label{eq:14}
\end{equation} 
The above reduced Chern--Simons term is similar to (\ref{eq:10}), but it
differs in detail, because the coefficient of ${\rm tr} A_r \Omega$ is not the
same as in (\ref{eq:10}).  The reason for this may be understood by
considering the special, but interesting case when $-\Omega$ is an element of
an
$SU(2)$ algebra, so that a gauge transformation on (\ref{eq:12})
puts those expressions into the form (\ref{eq:9})
\cite{ref:7}.  However, the relevant gauge function is singular (multivalued)
on the sphere (because it removes the ${\rm ctn} \theta$ singularity of
$A_\phi$),  and the Chern--Simons term is not invariant against this
transformation.  Tracking the change in
$NW$, exposes another contribution so that finally one finds
\begin{equation}
NW = N \int^\infty_0 dr {\rm tr} \Bigr\{ \epsilon^{mn} \Psi_m
(D\Psi)_n - 2A_r \Omega \Bigl\}
\label{eq:new17}
\end{equation} 
in agreement with (\ref{eq:10}).

Alternatively one may describe the situation as follows\cite{ref:8}:  The
Chern--Simons expression is defined intrinsically and gauge invariantly in
terms of the curvature $F_{\mu\nu}$,
\begin{equation}
NW = -\frac{N}{16\pi} \int_M d^4x \, \epsilon^{\alpha\beta\gamma\delta}
{\rm tr} (F_{\alpha\beta} F_{\gamma\delta})
\label{eq:17a}
\end{equation} 
where the integration is over a 4-manifold $M$ whose boundary $\partial M$ is
the 3-manifold relevant to the Chern--Simons formula.  When
connections are singular [like our (\ref{eq:12})] it need not be the case that
the ``surface'' integral in (\ref{eq:17a}) yields just (\ref{eq:8}) --- there
may be additional contributions.  We are interested in the Chern--Simons term
on
$R^3$, compactified to $S^3$.  With radial symmetry and identification of end
points in the $r$-coordinate, we pass to $S^1 \times S^2$.  So $M$ in
(\ref{eq:17a}) should be $D^2 \times S^2$, where $D^2$ is a disk with boundary
$S^1$.  Evidently, the extra contribution arises when this calculation is
carried out with (\ref{eq:12})\cite{ref:9}.

\section{Other Uses for the Dimensionally Reduced Chern--Simons Term}

The radially symmetric Chern--Simons term (\ref{eq:10}) or (\ref{eq:new17}) may
be inserted into the classical field equations of two theories.  One may
consider a Yang--Mills/Higgs model in $(3+1)$ dimensions, for which static 't
Hooft--Polyakov monopole solitons exist, with explicitly known profiles when
the Higgs potential vanishes (Bogomolny--Prasad--Sommerfield). 
Adding a 3-dimensional (spatial) Chern--Simons term violates Lorentz
invariance in an interesting, gauge-invariant fashion\cite{ref:10} and affects
the static, purely spatial equations of motion.

Alternatively, one may consider a $(2+1)$-dimensional Yang--Mills/Higgs theory,
continued to imaginary time, where the 't Hooft--Polyakov monopoles are now
instantons.  If the action in Minkowski space includes a Chern--Simons term,
continuation to imaginary time gives a Euclidean space action but with
imaginary Chern--Simons term.  Nevertheless the instanton equations remain
real.

In summary, one may sensibly add the Chern--Simons expression to the
3-dimensional Yang--Mills/Higgs system, with real or imaginary coefficient,
and inquire how the addition affects the monopole or instanton solutions,
respectively.  It appears that the inclusion of the Chern--Simons term removes
the 't Hooft--Polyakov monopoles/instantons in both contexts.  We sketch the
argument, for radially symmetric configurations in an
$SU(2)$ (Georgi--Glashow) theory.

The Yang--Mills/Higgs $SU(2)$ Lagrangian restricted to radially symmetric
and static gauge fields as in (\ref{eq:9}) and Higgs fields
\begin{equation}
\phi_a = \hat{r}^a f
\label{neweq:17}
\end{equation} 
takes the form (with kinetic term normalized to $1/32 \pi$ rather than the
usual $1/4$)
\begin{equation}
-L=\int^\infty_0 dr \, \Bigr\{ \frac12 (D\psi)_m (D\psi)_m + \frac{1}{4r^2}
(\psi^2_m -1)^2 + \frac12 \psi^2_m f^2 + \frac{r^2}{4} (f')^2 + V(f) \Bigl\}
\label{eq:15}
\end{equation} 
This corresponds to the static portion of the $(3+1)$-dimensional theory, with
$-L$ being also the energy.  Alternatively, (\ref{eq:15}) is the Euclidean
action of a $(2+1)$-dimensional theory continued to imaginary time.  The Higgs
potential $V(f)$ vanishes in the Bogomolny--Prasad--Sommerfield limit, which
we adopt henceforth.

To (\ref{eq:15}) we add the Chern--Simons term (\ref{eq:10}), arriving at
the total action
\begin{equation}
I=-L+(i) NW
\label{eq:16}
\end{equation} 
where the imaginary factor $i$ is present for the discussion of the
$(2+1)$-dimensional theory in imaginary time.  When the fields $\psi_m$ are
parametrized as
\begin{equation}
\psi_1=\rho \cos \theta \qquad \quad \psi_2 = \rho \sin \theta
\label{eq:17}
\end{equation} 
one is left with
\begin{eqnarray}
&&\hskip-2cm I=\int^\infty_0 dr \, \Bigr\{ \frac12 \rho'^2 + \frac12 \rho^2
(\theta' + A + (i)N)^2 -
\frac12 ((i)N)^2 \rho^2 + \frac{1}{4 r^2} (\rho^2 -1)^2  \nonumber \\
&&\qquad\qquad{}+ \frac12 \rho^2 f^2+
\frac{r^2}{4} f^{\prime 2} - (i) NA \Bigl\} \ .
\label{eq:18}
\end{eqnarray} 
The equation of motion obtained by varying $\theta$, 
\begin{equation}
\frac{d}{dr} \Bigl\{ \rho^2 (\theta' +A + (i)N) \Bigr\} = 0
\label{eq:19}
\end{equation} 
requires the constancy of the ``angular momentum'' $\rho^2(\theta'+A+(i)N)$,
but its value is unrestricted.  The stronger constraint follows from the
``Gauss law'' that emerges after varying~$A$.
\begin{equation}
\rho^2 (\theta'+A+(i)N) = (i)N
\label{eq:20}
\end{equation} 
The ``angular momentum'' is fixed.  In the equation for $\rho$,
$(\theta'+A+(i)N)$ may be eliminated with the help of (\ref{eq:20}), which
implies that an effective action for $\rho$ and $f$ reads
\begin{equation}
I_{\rm effective} = \frac12 \int dr \, \biggl\{ {\rho'}^2 \mp N^2 \Bigl(
\frac{1}{\rho^2} + \rho^2 \Bigr) + \frac{1}{2r^2} (\rho^2 -1)^2 + 
\rho^2f^2 + \frac{r^2}{2} f^{\prime 2} \biggr\}
\label{eq:21}
\end{equation} 
where the upper sign corresponds to the static soliton equation and the
lower to the imaginary-time instanton equation.  Note that the latter
is real, inspite of imaginary factors at intermediate stages.

It is seen that the addition of the Chern--Simons term modifies the equation
for $\rho$ by a ``centrifugal'' potential $\pm
N^2(\frac{1}{\rho^2} + \rho^2)$.  As long as $f$ remains bounded at large $r$,
this potential prevents $\rho$ from
achieving its monopole asympotote
$\rho=0$, in the monopole problem ($+$ sign).  For the instanton problem ($-$
sign), the potential is unboundedly attractive, so that the large $r$
asymptote corresponds either to unbounded growth of $\rho$, or to a collapse
to the origin at finite (yet large) $r$, or, in a limiting situation, to
$\rho$ attaining the maximum of the potential and remaining there.  But none
of the above corresponds to the instanton boundary condition.

Thus, the addition of the topological Chern--Simons interaction, destroys the
topological excitations.  This may be seen as an analog to the result that in
the vacuum sector of a $(3+1)$-dimensional theory with a (Lorentz
non-invariant) Chern--Simons interaction, an emergent mass term is
tachyonic\cite{ref:10}.


\newpage

\begin{thebibliography}{99~}

\baselineskip=20pt

\frenchspacing

\bibitem{ref:1} G.~Dunne, K.~Lee and C.~Lu, Phys.~Rev.~Lett. {\bf 78},
3434 (1997).

\bibitem{ref:2}S.~Deser, L. Griguolo and D.~Seminara, Phys. Rev. Lett. {\bf
79}, 1976 (1997) and preprint, hep-th/9712066; C. Fosco, G. Rossini and F.
Schaposnik, Phys. Rev. Lett. {\bf 79}, 1980 (1997) and Phys. Rev. D {\bf 56},
6547 (1997).

\bibitem{ref:3} The first of many calculations is by A. Niemi and G. Semenoff,
Phys. Rev. Lett. {\bf 51}, 2077 (1983).

\bibitem{ref:4}S.~Deser, R.~Jackiw and S.~Templeton, Phys. Rev. Lett. {\bf 48},
975 (1982), Ann. Phys. (NY) {\bf 140}, 372 (1982), (E) {\bf 185}, 406 (1988).

\bibitem{ref:5} A. Redlich, Phys. Rev. Lett. {\bf 52}, 18 (1984); Phys. Rev. D
{\bf 29}, 2366 (1984).

\bibitem{ref:6} G.~Dunne, R. Jackiw and C. Trugenberger, Phys. Rev. D {\bf
41}, 611 (1990); G.~Dunne and R. Jackiw, Nucl. Phys. B (Proc. Suppl.) {\bf
33C}, 114 (1993).

\bibitem{ref:7} The construction of symmetric connections is based on a
mathematical theorem due to H. Wong.  Physicists rediscovered and made
 this procedure explicit in 
P. Forg\'acs and N. Manton, Comm. Math. Phys. {\bf
72}, 15 (1980);
J. Harnad, S. Shnider and L. Vinet, J. Math. Phys. {\bf 21}, 2719 (1980).
For a detailed discussion of the radially symmetric configuration see
R. Jackiw, Acta Phys. Austr. Suppl. XXII, 383 (1980), reprinted in R. Jackiw,
{\it Diverse Topics in Theoretical and Mathematical Physics\/}, (World
Scientific, Singapore, 1995). 
Symmetry reduction of the Chern--Simons term has also been considered by E.
D'Hoker and L. Vinet (unpublished, private communication).

\bibitem{ref:8} We thank D. Seminara for discussion.

\bibitem{ref:9} V.P. Nair has
observed that group elements with odd-integer widing numbers are double-valued
on $S^1 \times S^2$, changing sign as $r$ interpolates between the identified
points $r=0$ and $r=\infty$.  If gauge transformations by such double-valued
group elements are excluded, then a half-integer quantization of the
one-dimensional Chern--Simons term would preserve gauge invariance of the
quantum theory. 


\bibitem{ref:10} S. Carroll, G. Field and R. Jackiw, Phys. Rev. D {\bf 41},
1231 (1990).

\nonfrenchspacing
\end{thebibliography}
\end{document}